\documentstyle[12pt,twoside,fleqn,espcrc1,epsf]{article}

\newcommand{\AmS}{{\protect\the\textfont2
  A\kern-.1667em\lower.5ex\hbox{M}\kern-.125emS}}

\title{Double $\Lambda$ Hypernuclei and the $\Lambda$-$\Lambda$
       Interaction.\thanks{
This research was supported by the DGES of Spain under 
contract PB95-1204 and by
the Junta de Andaluc\'\i a. J.C. acknowledges the spanish Ministry of 
Education for support through a postdoctoral fellowship.
}}

\author{
J. Caro\address{Physik Department, 
Technische Universit\"at M\"unchen, 
D-85747-Garching, Germany}
C. Garc\'{\i}a-Recio\address{
Departamento de F\'{\i}sica
Moderna, Universidad de Granada, 
E-18071 Granada, Spain}
        and
J. M. Nieves$^{\rm b}$
}

\begin{document}
\maketitle

The  $\Lambda$-$\Lambda$ effective interaction, in the channel
$L=S=0$, in the nuclear medium  is fitted to the available 
binding energies, $B_{\Lambda\Lambda}$, of double $\Lambda$
hypernuclei: $^{6}_{\Lambda\Lambda}$He, $^{10}_{\Lambda\Lambda}$Be and
$^{13}_{\Lambda\Lambda}$B. The mesonic decay of these hypernuclei is
also investigated. Finally, this effective interaction  is used to predict the
binding energies and mesonic decays widths  of heavier double
$\Lambda$ hypernuclei. 
 
We approximate the double $\Lambda$ hypernuclei by  systems composed 
by two interacting $\Lambda$'s moving in the mean field potential 
created by the nuclear cores.  The $\Lambda$-core potentials
are adjusted to reproduce the binding energies of the corresponding 
single $\Lambda$ hypernuclei~\protect\cite{l-core}.  
A $\sigma$-$\omega$ meson 
exchange potential~\protect\cite{ll} is used for the $\Lambda$-$\Lambda$
interaction. We apply both the Hartree-Fock (HF) and variational
approximations to compute $B_{\Lambda\Lambda}$. 

In our framework, the parameters of the $\Lambda$-$\Lambda$ 
interaction are the couplings of the $\sigma$
($g_{\sigma\Lambda\Lambda}$) and $\omega$ ($g_{\omega\Lambda\Lambda}$) mesons 
to the hyperon $\Lambda$ and the  cutoffs 
($\Lambda_{\sigma\Lambda\Lambda}$, $\Lambda_{\omega\Lambda\Lambda}$) of the
corresponding monopolar form-factors.
In ref.~\protect\cite{ll}, where N-N and N-Hyperon
low energy scattering data are analyzed in a similar framework, 
$g_{\omega \Lambda\Lambda}$  is fixed. 
There, it is also
found that apart from the $\sigma$-coupling, only the cutoffs at
vertices where vector mesons are involved are of sizeable
influence. Thus, we fix $\Lambda_{\omega\Lambda\Lambda}=2$ GeV, as
found in~\protect\cite{ll}. To check
the sensitivity of the hypernucleus data to
$g_{\omega\Lambda\Lambda}$ and  $\Lambda_{\sigma\Lambda\Lambda}$,
we consider different families of potentials obtained by combining 
different values of the ratio $g_{\omega\Lambda\Lambda}/g_{\omega
NN}$ (above and below the SU(3) prediction 2/3) and 
of the cutoff $\Lambda_{\sigma\Lambda\Lambda}$ (1 and 2 GeV). 
Thus, for each potential we are left
with just one free parameter ($g_{\sigma\Lambda\Lambda}$), 
which is obtained from a best $\chi^2$-fit to the values of  
$B_{\Lambda\Lambda}$ for the three discovered hypernuclei. 

\begin{figure}[htb]
\vspace{-1.5cm}
\leavevmode
\epsfysize = 600pt
\makebox[100mm]{\epsfbox{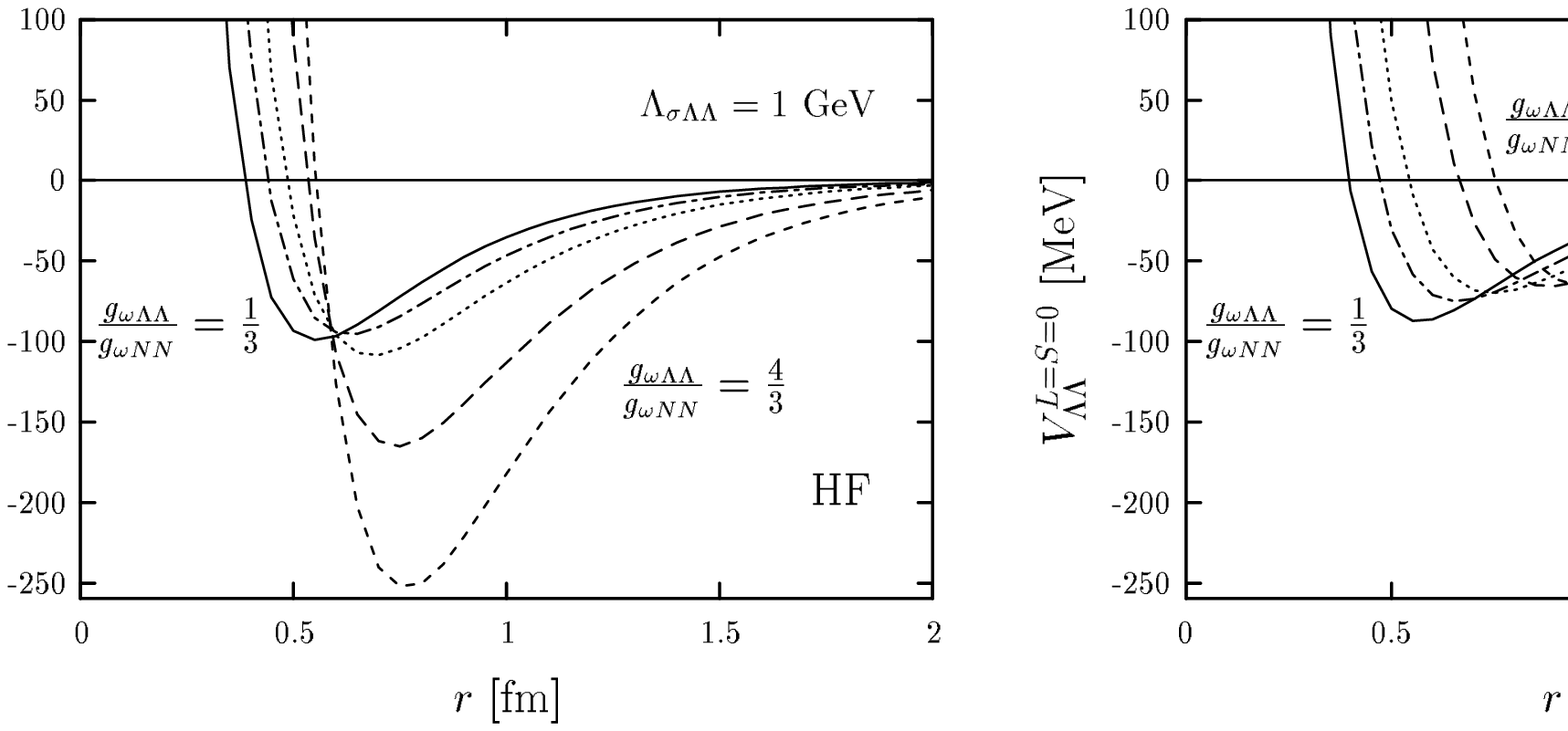}}
\vspace{-14.3cm}
\caption{Different $\Lambda$-$\Lambda$ potentials, 
in the channel $L=S=0$, obtained from fits
to $B_{\Lambda\Lambda}$ within the HF (left) and Variational
(right) approachs. In both
figures the five curves correspond to the values 1/3, 1/2, 2/3, 1 
and 4/3 for the ratio $g_{\omega\Lambda\Lambda}/g_{\omega NN}$. 
The cutoff mass $\Lambda_{\sigma\Lambda\Lambda}$ is 1 GeV 
} 
\label{fig:figure1}
\end{figure}

In Figure~\ref{fig:figure1} we show different $\Lambda$-$\Lambda$
potentials, obtained from the best fit to the data using the
HF approach (left) and the variational approach (right). 
Despite the very different shape and magnitude of
the interactions shown in the figure, all of them give values 
for $\chi^2$ per degree of freedom, $\chi^2/dof$, 
of the order of 0.1 and 0.5 for the HF and variational schemes
respectively, indicating the goodness of the
fits and the impossibility of selecting either of them only by means 
of the binding energies, $B_{\Lambda\Lambda}$. 
The variational approach lead to quite less atractive potentials than
the HF ones,
because of the great importance of the $|\vec{r}_1-\vec{r}_2|$-correlations 
included in the variational approach. The variational results together
with the detailed method of calculation will be described 
elsewhere.\cite{Ca98} Thus, the remaining results we present here have
been obtained in the HF approach.

From each fit we determine the parameter $g_{\sigma\Lambda\Lambda}$ 
with a statistical
error smaller than  $2\%$. 
To estimate the size of the systematic error we
consider different choices for the $\Lambda$-core potential, 
finding that in most cases it is of the order of the statistical one.
In ref.~\protect\cite{ll}, the $\Lambda$-p scattering data
were fitted with $g_{\omega\Lambda\Lambda}/g_{\omega NN}=$2/3 ,
$\Lambda_{\sigma\Lambda\Lambda}=1$ GeV and
$g_{\sigma\Lambda\Lambda}/\sqrt{4\pi }$=2.138.  
For this value of $g_{\omega\Lambda\Lambda}$ we
find $g_{\sigma\Lambda\Lambda}/\sqrt{4\pi}=3.11 \pm 0.04$
($\Lambda_{\sigma\Lambda\Lambda}=1$ GeV) or $2.28 \pm 0.02$
($\Lambda_{\sigma\Lambda\Lambda}=2$ GeV). Those couplings are
not directly comparable, because the ones determined in this work
correspond to an effective interaction in the nuclear medium. 

It is remarkable that for a fixed value of
$\Lambda_{\sigma\Lambda\Lambda}$, 
there is a linear correlation between $g^2_{\omega\Lambda\Lambda}$ 
and the fitted  $g^2_{\sigma\Lambda\Lambda}$. Indeed, 
\[ \frac{g^2_{\sigma\Lambda\Lambda}}{4\pi} = 
\left \{\begin{array}{lrr} (2.24 \pm 0.17) + (0.825 \pm 0.009) \times 
 g^2_{\omega\Lambda\Lambda}/{4\pi},\,\,\, & \chi^2/dof = 0.05,\,\,\, &
\Lambda_{\sigma\Lambda\Lambda} = 1 \,\,{\rm GeV} \\  
(2.21 \pm 0.18) + (0.836 \pm 0.010) 
\times g^2_{\omega\Lambda\Lambda}/{4\pi}, 
\,\,\,  & \chi^2/dof = 0.01,\,\,\, &
\Lambda_{\sigma\Lambda\Lambda} = 2 \,\,{\rm GeV} \end{array}\right. \]
Also a correlation is found between the
parameters $g_{\sigma\Lambda\Lambda}$ and
$\Lambda_{\sigma\Lambda\Lambda}$ so that the quantity
$g_{\sigma\Lambda\Lambda}\times 
\left (1-m^2_\sigma/\Lambda^2_{\sigma\Lambda\Lambda} \right ) $
remains constant within approximately $3\%$. 

\begin{figure}[htb]
\vspace{-1.7cm}
\leavevmode
\epsfysize = 750pt
\begin{center}
\makebox[100mm]{\epsfbox{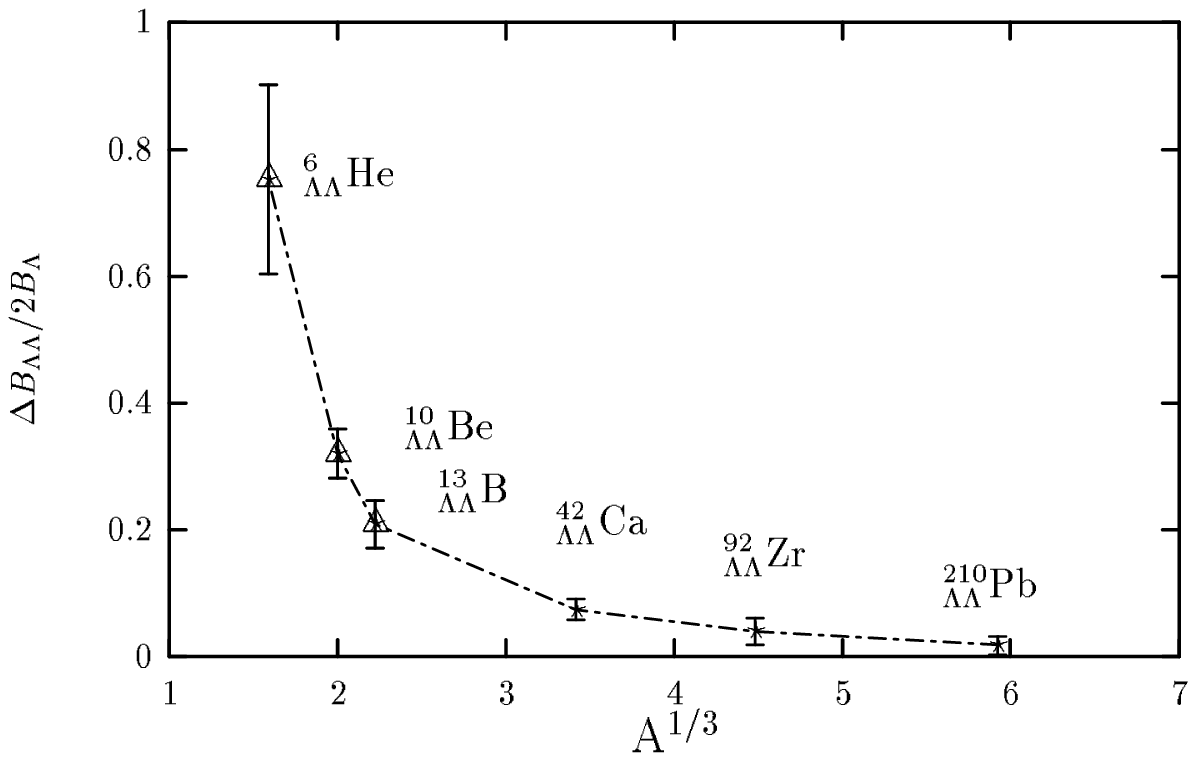}}
\end{center}
\vspace{-18.5cm}
\caption{ Binding energies of double $\Lambda$ Hypernuclei  
(triangles: experimental data, dotted line: theoretical calculation
including theoretical error bars).
}
\label{fig:figure2}
\end{figure}

We calculate the binding energies of heavier double hypernuclei 
($^{42}_{\Lambda\Lambda}$Ca, $^{92}_{\Lambda\Lambda}$Zr and  
$^{210}_{\Lambda\Lambda}$Pb) using the found family of $\Lambda$-$\Lambda$
potentials. They do not
depend much on the specific potential used. 
The results, shown in Figure~\ref{fig:figure2}, contain 
the statistical and systematic errors of the theoretical
calculations. 

\begin{figure}[htb]
\vspace{-1.7cm}
\leavevmode
\epsfysize = 750pt
\begin{center}
\makebox[100mm]{\epsfbox{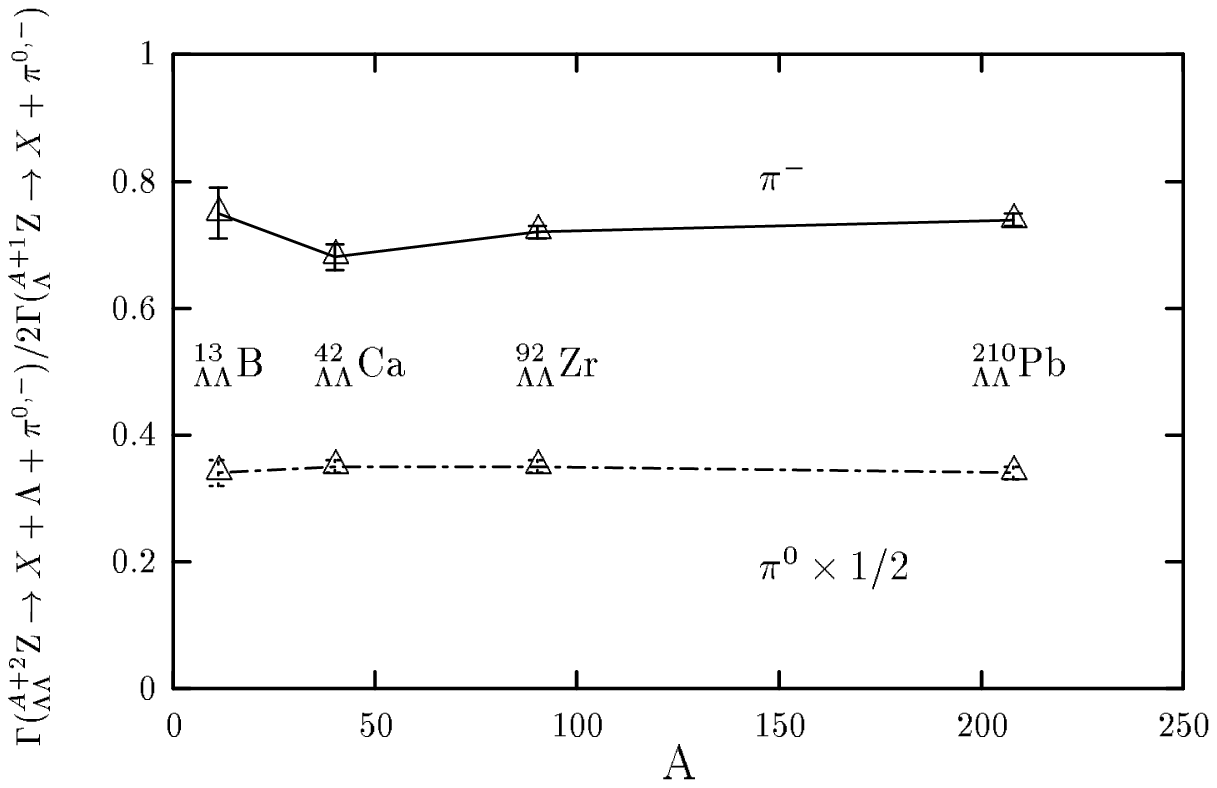}}
\end{center}
\vspace{-18.5cm}
\caption{ Calculated ratios of the mesonic decay widths of a $\Lambda$ to
$\pi^-$ (solid line) and to $\pi^0$ (dashdotted line,   shown with a
factor 1/2) for different hypernuclei using the HF approach.}
\label{fig:figure3}
\end{figure}

One might think that the mesonic decay of these double hypernuclei
could depend significantly on the details of the $\Lambda$-$\Lambda$
effective interaction and thus it could be used to differentiate
between the different potentials shown in
Figure~\ref{fig:figure1}. There are some theoretical uncertainties in
the calculation of the mesonic decay of
$^{5}_{\Lambda}$He~\protect\cite{l-core},\protect\cite{St93}, related to the nuclear core-$\Lambda$
interaction, and this has prevented us to look at $^{6}_{\Lambda\Lambda}$He
to check the dependence of the mesonic decay on the details of the 
$\Lambda$-$\Lambda$ interaction. Thus, we have looked at the
case of $^{13}_{\Lambda\Lambda}$B. The mesonic decay has been
computed following the method exposed in ref.~\protect\cite{St93}. Using  
the ten  different $\Lambda-\Lambda$ potentials obtained by the
best fit procedure with the HF approach (five of those potentials are
shown in Figure~\ref{fig:figure1} (left)),
we find that the mesonic decay width varies at most 
a $6\%$, making then this quantity inappropriate to
choose between the different potentials discussed above. However,
this fact allows us to make an accurate prediction of the mesonic
decay, and thus we find
\begin{eqnarray}
{\Gamma(^{13}_{\Lambda\Lambda}{\rm B} \to X + \Lambda + \pi^0
)}/
{\Gamma_\Lambda} & = & 0.076\pm 0.004, 
\nonumber\\ 
{\Gamma(^{13}_{\Lambda\Lambda}{\rm B} \to X + \Lambda + \pi^-
)}/
{\Gamma_\Lambda} & = & 0.37\pm 0.02,
\nonumber\end{eqnarray}
\noindent
where $\Gamma_\Lambda$ is the mesonic decay width of the
$\Lambda$ in the vacuum. 
The predicted results for some double hypernuclei of the ratios
$ {\Gamma(^{A+2}_{\Lambda\Lambda}{\rm Nucl} \to X + \Lambda + \pi^-
)}/
{2\Gamma_\Lambda(^{A+1}_{\Lambda}{\rm Nucl} \to X + \pi^-
)}$ and 
$ {\Gamma(^{A+2}_{\Lambda\Lambda}{\rm Nucl} \to X +\Lambda  + \pi^0
)}/{2\Gamma_\Lambda(^{A+1}_{\Lambda}{\rm Nucl} \to X + \pi^0
)}(\times 1/2)$  are shown in Figure~\ref{fig:figure3}.
The errors of those ratios are of the order of 5\%. 
They differ from the naively expected value 1, being around 0.7 for the
calculated hypernuclei.

\end{document}